\documentclass[aps,pra,superscriptaddress,preprint,amsmath,amssymb,showpacs]{revtex4-2}
\usepackage[english]{babel}
\usepackage[utf8]{inputenc}
\usepackage{amssymb,bm}
\usepackage{graphicx}
\usepackage{amsmath}
\usepackage{color}
\usepackage{comment}
\usepackage[colorlinks=true,citecolor=blue,urlcolor=blue]{hyperref}
\begin{document}
 \begin{titlepage}

\title{Discrete spectrum of edge states in a two-dimensional topological insulator}
\author{A.~I.~Milstein}\email{a.i.milstein@inp.nsk.su}
\affiliation{Budker Institute of Nuclear Physics of SB RAS, 630090 Novosibirsk, Russia}
\affiliation{Novosibirsk State University, 630090 Novosibirsk, Russia}
\author{I.~S.~Terekhov}\email{i.s.terekhov@gmail.com}
\affiliation{School of Physics and Engineering, ITMO University, 197101 St. Petersburg, Russia}

\date{\today}

\begin{abstract}

The interaction model of two electrons in the edge states of a two-dimensional topological insulator is investigated. Both solutions of the Schr\"odinger equation  and  solutions of the Bethe-Salpeter equation at different values of the Fermi energy are considered. It is shown that for the Bethe-Salpeter equation, which takes into account the existence of the Fermi surface, there is a discrete spectrum in the system of two electrons. This phenomenon does not appear in the case of the Schr\"odinger equation.
\end{abstract}

\maketitle
 \end{titlepage}
\section{Introduction}

Topological effects in solid state physics have been studied since the early 80s of the last century, see, e.g., Refs. \cite{Laughlin_1981, Halperin1982,Haldane1988}. Interest in such effects increased after the fabrication of graphene \cite{Novoselov2004} and investigation of the quantum spin Hall effect in it \cite{Kane_2005_1}.  It has been shown that edge states appear in graphene when the spin-orbit interaction is taken into account, and the spin of the edge states correlates  with the direction of their momentum. Similar states exist in two-dimensional topological insulators \cite{Hasan2010,Qi2011}.  In Refs. \cite{Bernevig2006_1,Bernevig2006_2} it is shown that there is a one-dimensional boundary between the ordinary and the topological phases in a two-dimensional topological insulator. At the boundary, one-dimensional states appear, and they can be described by the  Hamiltonian
\begin{eqnarray}\label{H1}
\hat{H}=v_F\hat{\sigma}_z \hat{p}.
\end{eqnarray}
Here the $z$-axes is directed along the boundary, $\hat{\sigma}_z$ is the Pauli matrix, $\hat{p}=-i\hbar \partial_z$ is the operator of momentum, $v_F$ is some constant that has the dimension of velocity. The states described by the Hamiltonian \eqref{H1} are called chiral. Similar Hamiltonians appear in the study of electronic properties of polymers, see the Su-Schrieffer-Heeger model \cite{SSH}. Note that in different systems the matrix $\hat{\sigma}_z$ corresponds to different degrees of freedom.  For example, in topological insulators it corresponds to the electron spin and in the Su-Schrieffer-Heeger model to the pseudospin. 

The transport properties of chiral states in topological insulators have been studied both theoretically and experimentally, see, e.g., \cite{Konig2007,Konig2013,Knez2014,Spanton2014,Tang2017,Fei2017,Cheianov2013,Vayrynen2017,
Kurilovich2019,Yevtushenko2022} and references therein. In the present paper, we study the interaction of two chiral electrons using two different approaches.  First, we study the interaction of two electrons without considering the Fermi surface. In this case, the Hamiltonian of the system is the sum of Hamiltonians for free chiral electrons and the potential interaction energy $V(z_1-z_2)$:
\begin{eqnarray}\label{SchredingerEq}
\hat{H}=\hat{\sigma}_1 \hat{p}_1+\hat{\sigma}_2 \hat{p}_2+V(z_1-z_2),
\end{eqnarray}
where $z_1$,  $z_2$ are the electron coordinates,  $\hat{p}_{1,2}=-i\frac{\partial}{\partial z_{1,2}}$ are the corresponding momentum operators, $V(z)=V(-z)$. Here and below we have omitted the index $z$ of the $\hat{\sigma}$-matrices and set $v_F=\hbar=1$.  The Schr\"odinger equation for the Hamiltonian \eqref{SchredingerEq} can be solved analytically. Second,to take into account the Fermi surface, we solve the Bethe-Salpeter equation. 

Let's make the following remark. It is known that the Landau Fermi-liquid theory is not applicable in the case of one spatial dimension. In addition, in the experimental paper \cite{Stuhler2020} it was shown that the edge states of  topological insulator are described by the Tomonaga-Luttinger liquid model rather than by the Landau Fermi-liquid theory. Therefore, it may seem that the Fermi surface is always absent in a one-dimensional system,  and its account using the Bethe-Salpeter equation is not adequate. However, the presence of two-dimensional states leads to a violation of bosonization in the one-dimensional electron gas in a topological insulator \cite{Hohenadler2012}. Therefore, if the number of filled chiral states in the band gap is small, then the Fermi surface exists. This means that the Bethe-Salpeter equation, which we use to describe two interacting electrons, is applicable.

In the present paper we  show that accounting of the Fermi surface in the case of one spatial dimension leads to qualitatively new effects that are absent for solutions of the Schr\"odinger equation. Namely, localized states arise that correspond to a discrete spectrum. Besides, the discrete spectrum depends on the total momentum of electrons.

\section{Schr\"odinger equation} 
Let us write the Schr\"odinger equation for the Hamiltonian \eqref{SchredingerEq} in the form:
\begin{eqnarray}
i\frac{\partial \psi}{\partial t}=-i\left[\left(\hat{\sigma}_1-\hat{\sigma}_2\right)\frac{\partial \psi}{\partial z}+\frac{\hat{\sigma}_1+\hat{\sigma}_2}{2}\frac{\partial \psi}{\partial Z} \right]+V(z)\psi, \label{EqPhi}
\end{eqnarray}
where $z=z_1-z_2$, $Z=(z_1+z_2)/2.$ Since this equation contains only $z$ components of the Pauli matrices, the projections of spins (pseudospins) of each electrons  on the $z$ axis  are conserved, and the operators $\hat{\sigma}_{1,2}$ can be replaced by the numbers $\sigma_{1,2}=\pm1$.  Therefore, we can write the wave function in the form:
\begin{eqnarray}\label{PhiRepr}
\psi(t,Z,z)&=&a(Z,z,t)|\!\!\uparrow\rangle_1|\!\!\uparrow\rangle_2+b(t,Z,z)|\!\!\downarrow\rangle_1|\!\!\downarrow\rangle_2\nonumber\\
&+&c(t,Z,z)|\!\!\uparrow\rangle_1|\!\!\downarrow\rangle_2+d(t,Z,z)|\!\!\downarrow\rangle_1|\!\!\uparrow\rangle_2\,.
\end{eqnarray}
Here $|\!\!\uparrow\rangle$ and  $|\!\!\downarrow\rangle$ are the eigenfunctions of the matrix $\hat{\sigma}$: $\hat{\sigma}_i|\!\!\uparrow\rangle_i=|\!\!\uparrow\rangle_i$, $\hat{\sigma}_i|\!\!\downarrow\rangle_i=-|\!\!\downarrow\rangle_i$. 
Substitution of the wave function in the form \eqref{PhiRepr} into Eq. \eqref{EqPhi} results in four independent equations.  For the functions $a$ and $c$, we have
\begin{eqnarray}
\frac{\partial a}{\partial t} +\frac{\partial a}{\partial Z}&=&-iV(z)a,\label{Eq_a0}\\
\frac{\partial c}{\partial t}+2\frac{\partial c}{\partial z} &=&-iV(z)c.\label{Eq_c0}
\end{eqnarray}
The equation for the function $b$ is obtained from Eq.~\eqref{Eq_a0} by replacing $Z\to -Z$, and the equation for $d$ is obtained from Eq.~\eqref{Eq_c0} by replacing $z\to -z$. Solving these equations by the method of characteristics, we get
\begin{eqnarray}
a(Z,z,t)&=&a_0(Z-t)a_1(z)e^{-iV(z)t},\label{a_Schr_Solution}
\end{eqnarray}
\begin{eqnarray}
c(Z,z,t)&=&c_0(z-2t)c_1(Z)\exp\left\{-\frac{i}{2}\int_0^zV(z')dz'\right\}\,.
\end{eqnarray} 
The functions $a_0$,  $a_1$,  $c_0$,  $c_1$ can be found from the conditions at $t=0$. It is seen that an account for the interaction potential does not lead to a nontrivial dynamics of electrons. Namely, up to the phase factor, the solutions at $t\neq0$ are obtained from the initial conditions by a shift of time in the arguments of corresponding functions. 

To interpret the wave function, we have derived the continuity equation from Eq.~\eqref{EqPhi}:
\begin{eqnarray}
\frac{\partial \rho}{\partial t}+ \frac{\partial}{\partial z} j_1+\frac{\partial}{\partial Z} j_2=0.
\end{eqnarray}
Here
$$\rho=|\psi(t,Z,z)|^2, \quad  j_1=\psi^+(\hat{\sigma}_1-\hat{\sigma}_2)\psi,\quad j_2=\frac{1}{2}\psi^+(\hat{\sigma}_1+\hat{\sigma}_2)\psi.$$  
It is seen that the current $j_1=0$ for the states with the spin structures $|\!\!\uparrow\rangle_1|\!\!\uparrow\rangle_2$ and $|\!\!\downarrow\rangle_1|\!\!\downarrow\rangle_2$. Also, the current $j_2=0$ for the states having the spin structures $|\!\!\uparrow\rangle_1|\!\!\downarrow\rangle_2$ and $|\!\!\downarrow\rangle_1|\!\!\uparrow\rangle_2.$ Therefore, the continuity equation splits into two independent equations:
\begin{eqnarray}
\frac{\partial \rho_{1}}{\partial t}+ \frac{\partial  j_1}{\partial z}=0\,,\\
\frac{\partial \rho_{2}}{\partial t}+ \frac{\partial j_2}{\partial Z}=0\,,\label{RhoSE}
\end{eqnarray}
\begin{eqnarray}
\rho_1&=&|c|^2+|d|^2,\quad j_1=2\left(|c|^2-|d|^2\right)\,,\\
\rho_2&=&|a|^2+|b|^2,\quad j_2= |a|^2-|b|^2\,.\label{Rho2J2Sch}
\end{eqnarray}
Since $\rho_1\ge 0$ and $\rho_2\ge 0$, these quantities can be interpreted as the densities of probabilities to find the electrons, having  antiparallel and parallel spins, respectively,  at the distance $z$ for a given $Z$.  Note that the current $j_1$ contains $|c|^2$ and $|d|^2$ with opposite signs due to the fact that the current changes its sign, $j_1\to -j_1$, at the replacement $z\to -z$. Similarly for the current $j_2$. Below we show that accounting for the Fermi surface leads to the properties of the system that are qualitatively different from those described above.

\section{Bethe-Salpeter equation}

In Ref.~\cite{MilstTer2019}, the influence of the Fermi surface on the interaction of electrons in graphene was considered by means of the Bethe-Salpeter equation. Using a similar approach for our case, we obtain the Bethe-Salpeter equation for a state having a certain energy $E$:
\begin{align}\label{BS_Eq0}
&[E-\sigma_1p_1-\sigma_2p_2]\Phi(p_1,p_2,E)=[\vartheta(\sigma_1p_1-E_F)+\vartheta(\sigma_2p_2-E_F)-1]
\nonumber\\
&\times\int_{-\infty}^{+\infty}\dfrac{dq}{2\pi}\,V_f(q)\,\Phi(p_1+q,p_2-q,E)\,,
\end{align}
where $V_f(q)$ is the Fourier transform of the potential $V(z)$, $\vartheta(x)$ is the Heaviside step function, $E_F$ is the Fermi energy, which is measured from the position of the Dirac point. This point corresponds to the energy of single electron with zero momentum. Let us represent the wave function $\Phi$ in the form similar to Eq. \eqref{PhiRepr}.
\begin{align}\label{dec1}
	&\Phi(p_1,p_2,E)=A(p_1,p_2,E)|\!\!\uparrow\rangle_1|\!\!\uparrow\rangle_2 +B(p_1,p_2,E)|\!\!\downarrow\rangle_1|\!\!\downarrow\rangle_2\nonumber\\
	&+C(p_1,p_2,E)|\!\!\uparrow\rangle_1|\!\!\downarrow\rangle_2+D(p_1,p_2,E)|\!\!\downarrow\rangle_1|\!\!\uparrow\rangle_2\,,
\end{align}
where the functions $A(p_1,p_2,E)$ and $C(p_1,p_2,E)$ satisfy the equations
\begin{align}\label{AC}
	&[E-p_1-p_2]A(p_1,p_2,E)=[\vartheta(p_1-E_F)+\vartheta(p_2-E_F)-1]
	\nonumber\\
	&\times\int_{-\infty}^{+\infty}\dfrac{dq}{2\pi}\,V_f(q)\,A(p_1+q,p_2-q,E)\,,\nonumber\\
	&[E-p_1+p_2]C(p_1,p_2,E)=[\vartheta(p_1-E_F)+\vartheta(-p_2-E_F)-1]
		\nonumber\\
		&\times\int_{-\infty}^{+\infty}\dfrac{dq}{2\pi}\,V_f(q)\,C(p_1+q,p_2-q,E)\,.
\end{align}
The equations for $B(p_1,p_2,E)$ and $D(p_1,p_2,E)$ are obtained from Eqs.~\eqref{AC} by replacing $p_{1,2}\to -p_{1,2}.$

It is convenient to pass from $p_{1,2}$ to the variables corresponding to the total and relative momenta of the electron system, i.e., $P=p_1+p_2$ and $p=(p_1-p_2)/2$.  The equations in the new variables have the form:
\begin{eqnarray}
	(E-P)\,\overline{A}(P,p,E)&=&-\text{sgn}(Q)\vartheta(|Q|-|p|) 
	\nonumber\\
	&&\times\int_{-\infty}^{+\infty}\dfrac{dq}{2\pi}\,V_f(p-q)\,\overline{A}(P,q,E)\,, \label{AP}\\
	(E-2p)\,\overline{C}(P,p,E)&=&[\vartheta(p-Q)+\vartheta(p-P-Q)-1]
	\nonumber\\
	&&\times\int_{-\infty}^{+\infty}\dfrac{dq}{2\pi}\,V_f(p-q)\,\overline{C}(P,q,E)\,,\label{CP}
\end{eqnarray}
where $Q=E_F-P/2$, $$\overline{A}(P,p,E)=A(p_1,p_2,E)\,,\quad \overline{C}(P,p,E)=C(p_1,p_2,E)\,.$$ The analysis shows that the solution of Eq.~\eqref{CP} does not contain a discrete spectrum, while Eq. \eqref{AP} does.  Therefore, below we  consider only Eq.~\eqref{AP}.

It is well known that the physical interpretation of the wave function, which is a solution of the Bethe-Salpeter equation, is a nontrivial problem. In our case, to interpret the function $\overline{A}(P,p,E)$, it is necessary to find the corresponding density and current that satisfy the continuity equation. For this purpose we follow Ref.~\cite{MilstTer2019} and introduce two operators and two corresponding functions:
\begin{eqnarray}\label{Lpm}
L_\pm (P,p)&=& \vartheta(\mp Q)\vartheta(|Q|-|p|)\,,\\
\overline{A}_{\pm}(P,p,E)&=&L_\pm(P,p) \overline{A}(P,p,E)\,.
\end{eqnarray}
Since $\overline{A}(P,p,E)\propto \vartheta(|Q|-|p|)$, see Eq.~\eqref{AP}, we have $$\overline{A}(P,p,E)=\overline{A}_{+}(P,p,E)+\overline{A}_{-}(P,p,E).$$ Then, Eq.~\eqref{AP} for the functions $\overline{A}_{\pm}$ takes the form: 
\begin{eqnarray}\label{LocStateEqP1}
(E-P)\overline{A}_{+}(P,p,E)&=&L_+ \int \frac{d q}{2\pi}V_f(p-q)\overline{A}(P,q,E), ,\\
(E-P)\overline{A}_{-}(P,p,E)&=&-L_- \int \frac{d q}{2\pi}V_f(p-q)\overline{A}(P,q,E), .
\end{eqnarray}
By performing the Fourier transform and the standard calculations, we obtain the continuity equation:
\begin{eqnarray}\label{RhoBS}
\left(\frac{\partial }{\partial t}+\frac{\partial }{\partial Z}\right)\varrho(Z,z,t)={\cal F}(Z,z,t),
\end{eqnarray}
where
\begin{eqnarray}
\varrho(Z,z,t)=|{\cal A}_+(Z,z,t)|^2-|{\cal A}_-(Z,z,t)|^2,\label{DensityBS}
\end{eqnarray}
\begin{eqnarray}
{\cal F}(Z,z,t)&=&2\,\text{Im}\int dZ'dz'[{\cal A}_+^*(Z,z,t){\cal L}_+(Z-Z',z-z')\nonumber\\
&+&{\cal A}_-^*(Z,z,t){\cal L}_-(Z-Z',z-z')]V(z')[{\cal A}_+(Z',z',t)+{\cal A}_-(Z',z',t)]\,,\label{F}
\end{eqnarray}
\begin{eqnarray}\label{AFur}
{\cal A}_{\pm}(Z,z,t)=\int\frac{dEdPdq}{(2\pi)^3}e^{-iEt}e^{iPZ+iqz}\overline{A}_{\pm}(P,p,E),
\end{eqnarray}
\begin{eqnarray}\label{F_Fur}
{\cal L}_{\pm}(Z-Z',z-z')=\int\frac{dPdq}{(2\pi)^2}e^{iP(Z-Z')+iq(z-z')} L_{\pm}(P,p).
\end{eqnarray}
The explicit form of the functions ${\cal L}_{\pm}(Z,z)$ is rather cumbersome. We do not present it here because it is not used in further calculations. The left-hand side of  Eq.~\eqref{RhoBS} is similar to that in Eq.~\eqref{RhoSE}, but the right-hand side  differs from zero and contains the source ${\cal F}$. Let us show that $\int dZdz {\cal F}(Z,z,t)=0$. To do this, we integrate both sides of Eq.~\eqref{F} over $Z$ and $z$, use the relations \eqref{AFur}, \eqref{F_Fur} and the properties $L_+^{2}(P,p)=L_+(P,p)$, $L_-^{2}(P,p)=L_-(P,p)$. Finally we have
\begin{eqnarray}
\int dZdz{\cal F}(Z,z,t)&=2\,\text{Im}\int dZ'dz'V(z')|{\cal A}_+(Z',z',t)+{\cal A}_-(Z',z',t)|^2=0\,.\label{FInt}
\end{eqnarray}
Therefore, if the function $\varrho(Z,z,t)$ decreases fast enough at $Z\to\pm\infty,$
we arrive at the conservation law
\begin{eqnarray}\label{ContEqInt}
    \frac{\partial }{\partial t}\int dZdz \varrho(Z,z,t)=0\,.
\end{eqnarray}
The interpretation of the function $\varrho(Z,z,t)$ differs from the interpretation of the function $\rho_2(Z,z,t)$ obtained for the Schr\"odinger equation. Note that the function $\varrho(Z,z,t)$ is not a positive quantity. Therefore, we  can not  interpret it as the density of probability. In addition, the continuity equation for $\varrho(Z,z,t)$ contains the source ${\cal F}(Z,z,t)$, which reflects the local change in the density of electrons having the energy less than $E_F$. The change occurs as a result of the interaction of these electrons with the pair of electrons having an energy greater than $E_F$. It follows from Eq.~\eqref{ContEqInt}  that the total charge of the entire system consisting of electrons above and below the Fermi surface is conserved at any point in time.  

Let us pass to consideration of the bound states of two electrons. In Refs.~\cite{MilstTer2019, Sabio2010, LeeMilstTer2012} the problem of two interacting electrons with non-zero total momentum $P$ has not been solved due to the complexity of solving the equations for the two-dimensional system. For our one-dimensional system, it is possible to express the solutions with $P\neq 0$ through the solution with zero total momentum. Therefore, let us first consider the case $P=0$.

\subsection{Solution at $P=0$}
For $P=0$ we set $p_1=-p_2=p$, then Eq.~\eqref{AP} takes the form:  
\begin{align}\label{ACP0}
	&E\,A_1(p,E)=-\text{sgn}(E_F)\vartheta(|E_F|-|p|) 
	\int_{-\infty}^{+\infty}\dfrac{dq}{2\pi}\,V_f(p-q)\,A_1(q,E)\,,
\end{align}
where  the notation $A_1(p,E)=A(0,p,E)$ is introduced. Since $A_1(p,E)\propto\vartheta(|E_F|-|p|)$, Eq. \eqref{ACP0} can be represented as
\begin{align}\label{ACP_New}
	&E\,A_1(p,E)=-\text{sgn}(E_F) 
	\int\limits_{-|E_F|}^{|E_F|}\dfrac{dq}{2\pi}\,V_f(p-q)\,A_1(q,E)\,,
\end{align}
where $A_1(p,E)$ is defined on the interval $|p|\le |E_F|.$ This equation is a homogeneous Fredholm equation of the second kind with a symmetric bounded kernel. Therefore Eq.~\eqref{ACP_New} has only the discrete spectrum $\{{E}_n\}.$ Thus, energy levels can be numerated such that $|{E}_1|>|{E}_2|>|{E}_3|>\ldots$, and $|E_1|$ is bounded from above. The eigenfunctions corresponding to different eigenvalues are orthogonal. Note that the wave functions in the coordinate representation are localized with respect to the variable $z$. All of the above statements are valid for any regular potential $V(z)$.

Let us consider the model potential
\begin{eqnarray}
V(z)=u_0 \exp\left\{-z^2/R^2\right\}\,,
\end{eqnarray}
where $u_0$ is the constant having dimension of energy, $R$ is the characteristic size of the potential. The Fourier transform of this potential reads
\begin{eqnarray}
V_f(q)=\sqrt{\pi}u_0 R \exp\left\{-q^2R^2/4\right\}\,.\label{Pot_In_q}
\end{eqnarray}
Substituting  Eq.~\eqref{Pot_In_q} into Eq.~\eqref{ACP_New}, we obtain
\begin{eqnarray}\label{LocStateEq}
{E}{A}_1({p},{E})&=&-\sqrt{\pi}\,\text{sgn}(E_F u_0)\int\limits_{-|E_F|}^{|E_F|} \frac{d {q}}{2\pi}\exp\left\{-\frac{({p}-{q})^2}{4}\right\}{A}_1({q},{E})\,.
\end{eqnarray}
Here and below we measure the energy $E$ in units of $|u_0|$, the momenta $p$, $q$ and the energy $E_F$ in units of $1/R$. The solution of Eq.~\eqref{LocStateEq} can be found  numerically for arbitrary values of $E_F$ and analytically for $|E_F| \ll1.$  In the latter case we can use the expansion of the kernel in Eq.~\eqref{LocStateEq} at small values of ${p}$ and ${q}$. The first two eigenvalues and the corresponding eigenfunctions have the form:
\begin{eqnarray}
{E}_1&=&-\frac{|E_F|}{\sqrt{\pi}}\text{sgn}(u_0\,E_F)\,, \,\, A_1(p,E_1)=\sqrt{\frac{\pi}{ |E_F|}}\theta(|E_F|-|p|)\,,\\ 
E_2&=&-\frac{|E_F|^3}{6\sqrt{\pi}}\text{sgn}(u_0\,E_F)\,, \,\, A_1(p,{E}_2)=p\sqrt{\frac{3\pi}{|E_F|^2}}\theta(|E_F|-|p|)\,.
\end{eqnarray}
These functions in the coordinate representation are
\begin{eqnarray}
 {\cal A}_1(z,E_1)&=&\sqrt{\frac{|E_F|}{\pi}}\,\frac{\sin y}{y}\,,\\
 {\cal A}_1(z,E_2)&=&\sqrt{\frac{3|E_F|}{\pi}}\,\frac{\sin y-y\cos y }{y^2}\,,
\end{eqnarray}
where $y=|E_F| z$. The functions ${\cal A}_1(z,E_1)$ and ${\cal A}_1(z,E_2)$ are localized and have a characteristic scale $|z|\sim 1/|E_F|.$ Note that the spectrum is discrete for any sign of the parameters $u_0$ and $E_F$.

\begin{figure}
\includegraphics[width=0.5\linewidth,clip]{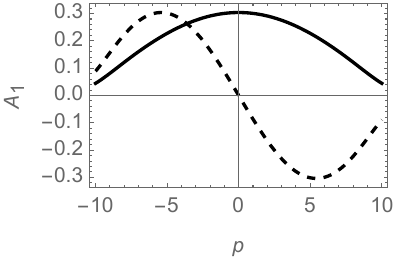}
\caption{The functions ${A}_1({p},E_{1,2})$ for $|E_F|=10$. Solid and dashed lines correspond to the functions ${A}_1({p},E_{1})$ and ${A}_1({p},E_{2})$, respectively. }\label{Function1_2p}
\end{figure}

For $|E_F| \gg 1$,  the energy levels obey the condition $$1\gg\frac{|E_{n}|-|E_{n+1}|}{|E_{n}|}>0\,.$$ In the leading order in the parameter $1/|E_F|$, we have ${E}_1=-\text{sgn}(u_0E_F).$  The wave functions ${A}_1({p},E_1)$ and ${A}_1({p},E_2)$, corresponding to the energies ${E}_1=-0.98\times\text{sgn}(u_0 E_F)$ and ${E}_2=-0.92\times\text{sgn}(u_0 E_F)$, are found numerically for $|E_F|=10$ and shown in Fig.~\ref{Function1_2p}. 

For $E_F<0$ and $u_0>0$ (the repulsive potential),  the energies are $E_i\ge0.$ For $E_F<0$ the projector $L_-(0,p)=0$, see \eqref{Lpm}, and ${L}_-(0,p)A_1(p,E_i)=0$. This means that the plane-wave decomposition of the wave function contains the states only with energies greater than the Fermi energy.  Therefore, such solutions correspond to the bound state of two electrons.

For $E_F<0$ and $u_0<0$ (the attractive potential),  the energies are $E_i\le0.$ Note that for a given $E_F<0$, there is such $u_0$ that $E_1<E_F$. This means that an energy level corresponding to the localized state can appear below the Fermi energy. Since the relation ${L}_-(0,p)A_1(p,E_i)=0$ is still satisfied, the solutions $A_1(p,E_i)$ correspond to bound states of two electrons. This is similar to the appearance of a Cooper pair in a superconductor.

When $E_F>0$, the wave functions satisfy the condition ${L}_+(0,p) A_1(p,E_i)=0.$ Therefore, if all electronic states having energies less than $E_F$ are filled, the solutions $A_1(p,E_i)$ are meaningless. 

The function $A_1(p,E)$ in Eq.~\eqref{ACP0} has the following property. If $A_1(p,E)$ is the solution, then $A_1(-p,E)$ is also the solution of Eq.~\eqref{ACP0}. This means that all solutions at a given energy are either symmetric or antisymmetric. It reflects the existence of only the discrete spectrum in our problem. The function $A_1(p,E)$, corresponds to the spin part $|\!\!\uparrow\rangle_1|\!\!\uparrow\rangle_2$ of the wave function. There is also the function $B_1(p,E)$ corresponding to the spin part $|\!\!\downarrow\rangle_1 |\!\!\downarrow\rangle_2$, we recall $B_1(p,E)=A_1(-p,E)$. Due to Pauli exclusion principle, the total wave function of two electrons must be antisymmetric at permutation of all coordinates and quantum numbers describing the state. Since the spin (pseudospin) part of the considered wave function is symmetric, only antisymmetric solutions $A_1(p,E)$ are realized, if other discrete degrees of freedom are absent. 

\subsection{Solution at $P\neq0$}
If we replace $E\to E-P$ and $E_F\to E_F-P/2=Q$ in Eq.~\eqref{ACP0} for the function $A_1(p,E)$, we arrive at Eq.~\eqref{AP} for the function $\overline{A}(P,p,E)$.  Therefore, the spectrum of Eq.~\eqref{AP} can be obtained from the spectrum of Eq.~\eqref{ACP0} using the above substitution. That is, let ${\cal E}_i(P,E_F)$ be the eigenvalue of Eq.~\eqref{AP} and let $E_i(E_F)$ be the eigenvalue of Eq.~\eqref{ACP0}, then
\begin{eqnarray}\label{Spectr}
{\cal E}_i(P,E_F)=P+E_i(E_F+P/2).
\end{eqnarray}
Thus, ${\cal E}_i(P,E_F)$ depends non-trivially on $P$.

\section{Conclusion}
The interaction between two electrons in the edge states of two-dimensional topological insulator is considered. The influence of the Fermi surface on this interaction is studied using the Bethe-Salpeter equation. It is shown that for an arbitrary potential $V(z)$ the solutions corresponding to parallel spins (pseudospins) contain only a discrete spectrum. An explicit relation \eqref{Spectr} between the spectra of the system of interacting electrons with total zero and nonzero momenta $P$ is obtained. The properties of explicit results, obtained for the model potential, are in agreement with the general conclusions. For the Schr\"odinger equation, which is valid for $E_F\to-\infty$, the discrete spectrum does not appear.

\section*{ACKNOWLEDGMENT}
The work of I.S.T. was financially supported by the ITMO Fellowship Program.

\end{document}